\documentclass{aa} 

\usepackage{natbib}
\bibpunct{(}{)}{;}{a}{}{,} 

\usepackage{graphicx}
\usepackage{txfonts}
\usepackage[dvipsnames]{xcolor}

\newcommand{\Fermi}{\textit{Fermi}\xspace}
\newcommand{\TXS}{TXS 0506+056\xspace}
\newcommand{\GB}{GB6 J1040+0617\xspace}
\newcommand{\PKS}{PKS 1502+106\xspace}
\newcommand{\nuPKS}{IceCube-190730A\xspace}

\begin{document}

   \title{\textit{Fermi}-LAT follow-up observations in seven years of real-time high-energy neutrino alerts}

   \author{S. Garrappa
          \inst{1,2}
           \and S. Buson \inst{3} \and J. Sinapius \inst{4} \and
          A. Franckowiak\inst{2} \and I. Liodakis \inst{5} \and C. Bartolini \inst{6,7} \and M. Giroletti \inst{8} \and C. Nanci \inst{8,9} \and G. Principe \inst{10,8} \and T. M. Venters \inst{11}
          }
   \institute{
   Department of Particle Physics and Astrophysics, Weizmann Institute of Science, 76100 Rehovot, Israel.
\and
   Fakultät für Physik \& Astronomie, Astronomisches Institut (AIRUB), Ruhr-Universität Bochum, D-44780 Bochum, Germany.\\
              \email{simone.garrappa@gmail.com} 
        \and
            Institut f\"{u}r Th. Physik und Astrophysik, University of W\"{u}rzburg,  Emil-Fischer-Str. 31 D-97074, W\"{u}rzburg, Germany
        \and
             Deutsches Elektronen-Synchrotron DESY, Platanenallee 6, 157387 Zeuthen, Germany.
        \and
             Finnish Centre for Astronomy with ESO, 20014 University of Turku, Finland.
        \and 
             Dipartimento di Fisica, Università di Trento, 38123 Trento, Italy.
        \and 
             INFN Sezione di Bari, 70125 Bari, Italy.
        \and 
             Istituto Nazionale di Astrofisica, Istituto di Radioastronomia (IRA), via Gobetti 101, 40129 Bologna, Italy.
        \and 
             Dipartimento di Fisica e Astronomia, Università di Bologna, via Gobetti 93/2, 40129 Bologna, Italy.
        \and Dipartimento di Fisica, Università di Trieste, I-34127 Trieste, Italy
             INFN, Sezione di Trieste, I-34127 Trieste, Italy.
        \and Astrophysics Science Division, NASA Goddard Space Flight Center, Greenbelt, MD, USA 20771.
             }

   \date{Published on Astronomy \& Astrophysics.}

 
  \abstract{
  The realtime program for high-energy neutrino track events detected by the IceCube South Pole Neutrino Observatory releases alerts to the astronomical community with the goal of identifying electromagnetic counterparts to astrophysical neutrinos. Gamma-ray observations from the  \textit{Fermi}-Large Area Telescope (LAT) enabled the identification of the flaring gamma-ray blazar \TXS as a likely counterpart to the neutrino event IC-170922A. By continuously monitoring the gamma-ray sky, \Fermi-LAT plays a key role in the identification of candidate counterparts to realtime neutrino alerts. In this paper, we present the \Fermi-LAT strategy for following up high-energy neutrino alerts applied to seven years of IceCube data. Right after receiving an alert, a search is performed in order to identify gamma-ray activity from known and newly detected sources that are positionally consistent with the neutrino localization. 
In this work, we study the population of blazars found in coincidence with high-energy neutrinos and compare them to the full population of gamma-ray blazars detected by \Fermi-LAT. We also evaluate the relationship between the neutrino and gamma-ray luminosities, finding different trends between the two blazar classes BL Lacs and flat-spectrum radio quasars.}


\maketitle


\section{Introduction}
Multimessenger astronomy has entered a new era with the identification of several candidate neutrino sources. Thanks to the improvements of neutrino telescopes and the multiwavelength programs of the astronomical community, dedicated observing campaigns have targeted promising candidates of various types, such as blazars \citep{IceCube:2018dnn}, nearby active galaxies \citep{2022Sci...378..538I} and tidal disruption events (TDE; \citealt{2021NatAs...5..510S}, \citealt{2022PhRvL.128v1101R}). A major boost in the association of astrophysical sources with high-energy (>100 TeV) neutrinos was given by the introduction of a realtime program of alerts by the IceCube South Pole Neutrino Observatory in 2016 \citep{2017APh....92...30A}. Since its detection in 2013, the origins of the diffuse high-energy neutrino flux \citep{2013Sci...342E...1I} remains one of the most compelling enigmas to solve.

High-energy neutrinos are produced in the decay of charged pions originated in the interactions of high-energy protons with matter (\emph{pp}) or with photon fields (\emph{p}$\gamma$). Neutral pions ($\pi^{0}$) that are produced alongside the charged pions decay into two gamma rays. Therefore, gamma rays are considered a promising tracer of hadronic interactions in astrophysical environments. However, gamma rays can also be produced in leptonic processes by means of bremsstrahlung or inverse Compton scattering.

Major challenges in the identification of these sources come from the limited angular resolution of neutrino detectors and from the uneven coverage of the sky of the majority of observing facilities. In addition, observational signatures for the identification of hadronic interactions in the spectra of sources are still highly debated, and interpretations of observed data with models that include photohadronic ($p\gamma$) and hadronuclear ($pp$) processes are affected by the often poor data coverage (see \citealt{2023ecnp.book..483M} for a review).
\newline
\newline
Observations from the Large Area Telescope (LAT) on board of the \Fermi Gamma-ray Space Telescope were crucial for the identification of the most statistically significant coincidence between a high-energy neutrino and an astrophysical source observed to date. On 22 September, 2017 a $\sim 290$ TeV neutrino event with a high probability of being astrophysical was detected in spatial coincidence with the gamma-ray blazar TXS 0506+056 \citep{IceCube:2018dnn}. Prompt follow-up observations by the \textit{Fermi}-LAT Collaboration \citep{Tanaka:2017atel} identified the blazar in an exceptional flaring state and triggered a rich multiwavelength campaign that involved 18 observational facilities from radio observatories up to imaging air Cherenkov telescopes. This collective effort unveiled the first compelling evidence for a blazar as a neutrino source.

Blazars, those active galactic nuclei (AGN) powered by accretion onto supermassive (M > 10$^{6}$ $M_{\odot}$) black holes that have their relativistic jets pointing at small angles towards the observer, have been proposed as prime candidates of extragalactic neutrino sources in several works \citep[e.g.][]{1993A&A...269...67M,2001PhRvL..87v1102A,dermer09,2012ApJ...749...63M,2015MNRAS.448..910C,2018ApJ...854...54R,2023ecnp.book..483M}. Although recent results of stacking analyses constrain the contribution of the gamma-ray blazar population to the whole diffuse neutrino flux \citep{Aartsen:2016lir,2023ApJ...954...75A}, several individual objects have been identified as possible neutrino sources. Before the introduction of the realtime program, a first neutrino blazar candidate was proposed from a coincidence between an IceCube cascade-type event and the blazar PKS 1424-24 which was flaring in radio and gamma rays \citep{2016NatPh..12..807K}. Another candidate counterpart to an archival high-energy neutrino event is the blazar \GB, which was flaring in the \Fermi-LAT and optical bands at the time of the detection of IceCube-141209A \citep{2019ApJ...880..103G}. After IceCube began issuing realtime alerts, another candidate neutrino source, the gamma-ray blazar \PKS was identified through its spatial coincidence with the event \nuPKS \citep{2019ATel12972....1G}. Although the source was not flaring in gamma rays at the time of the detection, multiwavelength observations and modeling provided additional evidence in support of it being considered a plausible neutrino emitter \citep{2020ApJ...893..162F,2021ApJ...912...54R,2021JCAP...10..082O}. \cite{2020ApJ...902..133K} shows how  observed fluences of individual blazar flares from keV to GeV energies are too small to yield significant detection ($\geq$ 1 event) probabilities with the current sensitivity of neutrino telescopes, consistent with the low number of candidate counterparts observed so far. Similar conclusions were drawn in \cite{2018A&A...620A.174K} by studying the average GeV properties of blazars coincident with high-energy neutrinos; although, recent theoretical studies on a broad sample of \Fermi-LAT blazars have predicted a promising correlation between gamma-ray and neutrino luminosities \citep{2024A&A...681A.119R}.   

Correlations between the blazar population and high-energy neutrinos are recently suggested from a variety of multiwavelength searches, indicating that gamma rays might not be the only tracer to identify a population of neutrino blazars. In \cite{2022ApJ...933L..43B} and \cite{2023arXiv230511263B} a sample of candidate PeVatron blazars was suggested based on a cross-correlation between the 5th Roma-BZCat catalog \citep{2015Ap&SS.357...75M} and the hot spots \citep{2017ApJ...835..151A,2022Sci...378..538I} from the IceCube all-sky analyses \citep[see also][]{bellenghi:2023}. Studies that focus on observations conducted in the radio band suggest correlations between bright radio blazars and high-energy neutrinos \citep{2020ApJ...894..101P,2021A&A...650A..83H,2023MNRAS.523.1799P}. However, these correlations are still not confirmed by recent analyses that use a richer set of IceCube data \citep{2023ApJ...954...75A}.

Recent studies of the broader population of AGN have shown that their cores contribute $>$27$\%$ of the observed diffuse neutrino flux \citep{2022PhRvD.106b2005A}. Evidence of $>$TeV neutrino emission has been found at a 4.2$\sigma$ significance from the nearby AGN NGC 1068, making it the brightest neutrino point source detected in the Northern Sky \citep{2022Sci...378..538I}.

In addition to AGN, TDEs have also garnered interest as possible neutrino emitters due to the identification of three candidate sources that were found to coincide with realtime IceCube alerts \citep{2021NatAs...5..510S,2022PhRvL.128v1101R,2024MNRAS.529.2559V}. However, none of these sources were significantly detected in \Fermi-LAT observations, leading to constraints on lepto-hadronic models (e.g. \citealt{2023ApJ...948...42W}).

In this paper, we present the strategy employed by the \Fermi-LAT team to follow up realtime neutrino alerts (Section~\ref{sec:fup_strategy}). 
In Section \ref{sec:fup_res}, we present the results from realtime and archival gamma-ray counterpart searches, including a discussion of peculiar coincidences with \Fermi-LAT catalog sources and newly detected emitters (Section \ref{sec:multiplet}). In Section \ref{sec:pks0735}, we present the results from the observations of the blazar PKS 0735+178, an interesting candidate counterpart to three high-energy neutrinos. In Section \ref{sec:gamma_neutrino_connection}, we present the results from a population study of candidate neutrino blazars. In Section \ref{sec:results} and \ref{sec:conclusions} we discuss our results and the future prospects for the identification of gamma-ray counterparts to high-energy neutrinos.

\section{Follow-up observations with \textit{Fermi}-LAT}
\label{sec:fup_strategy}
The LAT instrument is sensitive to gamma rays with energies from 20 MeV to greater than 300 GeV \citep{2009ApJ...697.1071A}. In the 15 years since its launch, it has been operating in survey mode, providing continuous monitoring of the entire gamma-ray sky.

Since the introduction of the IceCube Realtime Alert Stream in 2016, we have defined a follow-up strategy for neutrino alerts. It consists of a systematic analysis of the sky region around the neutrino arrival direction, looking for known sources that may be flaring or new gamma-ray emitters. The timescales of interest range from a single day up to the full set of historical observations obtained by the LAT over the entire mission. 

During a standard follow-up analysis using LAT data, we investigate three different timescales (defined relative to the neutrino arrival time, T$_{0}$):
\begin{itemize}
    \item 1 day before T$_{0}$: sensitive to the detection of fast, bright transients, down to a few-hours duration.
    \item 1 month before T$_{0}$: sensitive to the detection of recent transient or variable behavior from the sources of interest.
    \item Full-mission data up to T$_{0}$: study long-term behavior of LAT catalog sources and detect weak gamma-ray emitters not reported by the LAT catalogs.
\end{itemize}

The choice of the aforementioned timescales is motivated by a trade-off between the instrument sensitivity and the expected time lag between gamma-ray and neutrino emission from time-dependent studies of blazars (see \citealt{2019NatAs...3...88G} for an application to the gamma-ray flare of TXS 0506+056 in 2017).

After receiving the first GCN Notice with the preliminary localization of the neutrino event from the Astrophysical Multimessenger Observatory Network (AMON)/GCN stream, we perform the first checks on the available data in the region at the 1-day and 1-month timescales. After typically a few hours, when the refined neutrino localization is released via GCN Circulars by IceCube, the analysis is repeated for the new position at all standard timescales with data up to the neutrino arrival time T$_{0}$. This delay between the two GCN revisions is comparable to the typical delay needed by the Fermi Science Support Center (FSSC) servers to have the most recent data up to T$_{0}$ available.

When significant detections are found at short timescales within the 90\% neutrino uncertainty region, a light curve analysis starting from one year prior to T$_{0}$ is performed to characterize the temporal evolution of the source in the short and medium term. In the case of no significant detection at the neutrino best-fit position, we report 95$\%$ confidence level upper limits corresponding to the detection of a point source with a power-law spectrum with index 2.0.

For typical configurations of the LAT follow-up analysis, we select photons from the event class developed for point-source analyses in the energy range from 100 MeV up to 1 TeV binned into 10 logarithmically spaced energy intervals per decade. We select a region of interest (ROI) of at least 15$\times$15 degrees centered on the neutrino best-fit position (or larger for neutrino alerts with wide error contours) binned in 0.1 deg size pixels. The binning is applied in celestial coordinates using a Hammer–Aitoff projection. We use standard data-quality cuts to select alerts observed when the detector was in a normal operation mode and remove time periods with the Sun within 15 deg of the ROI center. We perform a maximum-likelihood analysis using the latest available version of the \textit{Fermi}-LAT ScienceTools package\footnote{\url{https://fermi.gsfc.nasa.gov/ssc/data/analysis/software/}} (along with the latest IRFs and diffuse models) from the FSSC and the Fermipy package \footnote{\url{https://fermipy.readthedocs.io/en/latest/}} \citep{2017ICRC...35..824W}.

\begin{figure*} 
\sidecaption
    \includegraphics[width=12cm]{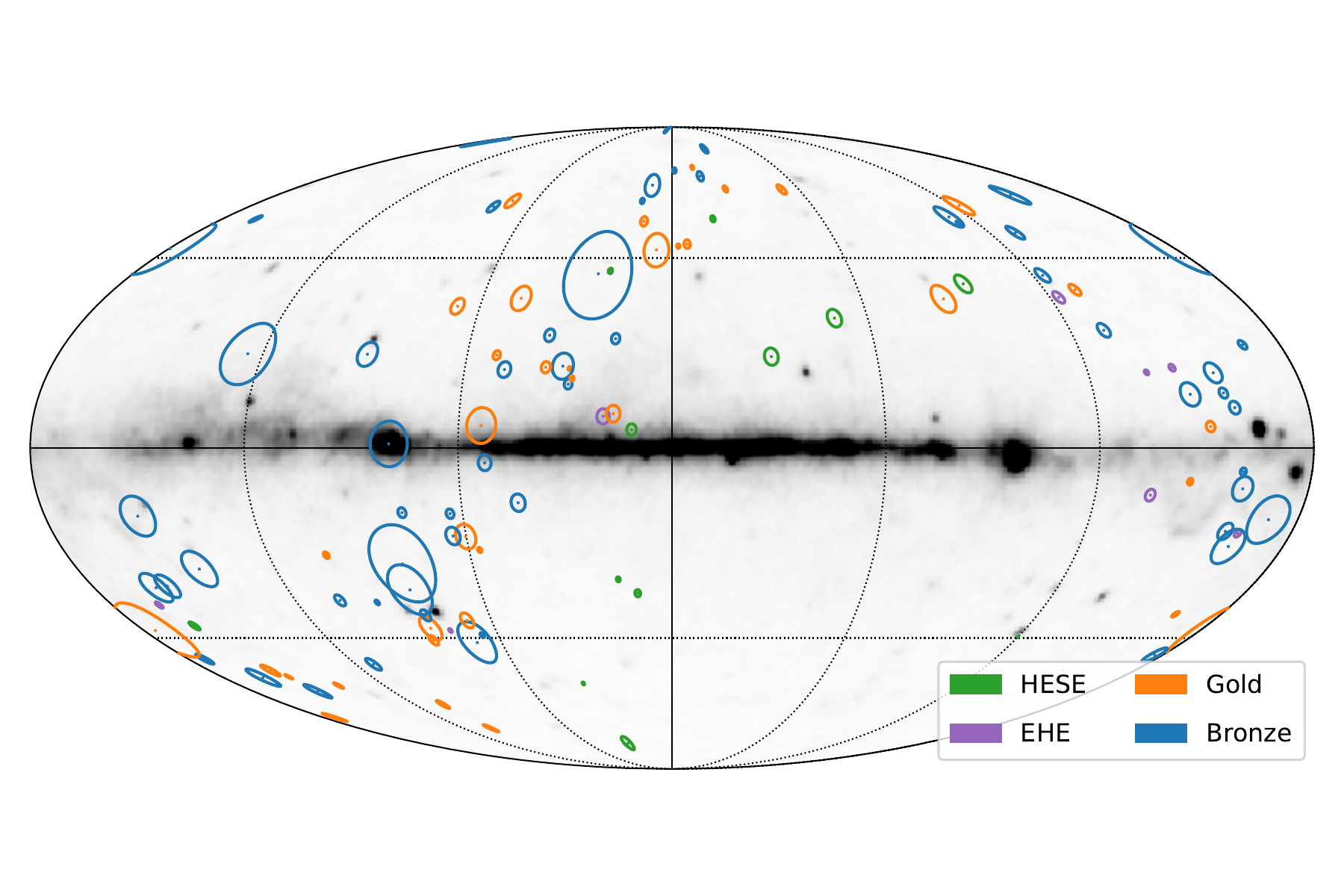}
        \caption{All-sky map showing the best-fit positions and 90$\%$ containment regions (approximated as circles with same areas) of the IceCube Realtime Alert stream in galactic coordinates. Gold alerts are shown in orange, Bronze alerts in blue, HESE alerts in green and EHE in purple (see text for the definitions of the various alert classes). The underlying map in gray scale shows the photon counts in each Healpix pixel from one year of \Fermi-LAT data.}
        \label{fig:skymap}
\end{figure*}

\section{Follow-up observations and results}\label{sec:fup_res}
\subsection{IceCube Realtime Alert System 1.0}
\label{sec:fup_res1}
In this section, we consider neutrino track alerts issued between the start of the Realtime System 1.0 in 2016 up to May 31, 2019. The \textit{Fermi}-LAT observed all 21 alerts issued via GCN Notices by AMON. In the first program, realtime neutrino alerts were issued via one of two streams - high-energy starting events (HESE) or through-going, extreme high-energy events (EHE) - with neutrino events being classified according to their topology \citep{2017APh....92...30A}. In the three years of this program, 12 alerts were classified as HESE and 9 as EHE. The all-sky map in Figure \ref{fig:skymap} shows the best-fit localizations and 90$\%$ error contours of these alerts with HESE events in green and EHE events in purple.

 The left panel of Figure \ref{fig:errors_4fgl} shows the distributions of the neutrino 90$\%$ containment regions for each alert classification, ranging from 0.44 deg$^{2}$ up to $\sim$24 deg$^{2}$. The median extension of the full sample of neutrino 90$\%$ containment regions is 3.2 deg$^{2}$ (black dashed line) while it is 1.96 deg$^{2}$ for the EHE sample and 5.6 deg$^{2}$ for the HESE.
 
For 13 of the aforementioned 21 alerts, there are no coincident sources from the third data release of the Fourth \Fermi-LAT Gamma-ray Source Catalog (4FGL-DR3; \cite{2022ApJS..260...53A}). Three of the alerts are each coincident with a single 4FGL-DR3 candidate which is classified as a blazar. The remaining five alerts are coincident with several 4FGL-DR3 sources.\\

\subsection{IceCube Realtime Alert System 2.0}
\label{sec:fup_res2}
\begin{figure*}[h!]
    \includegraphics[width=0.5\linewidth]{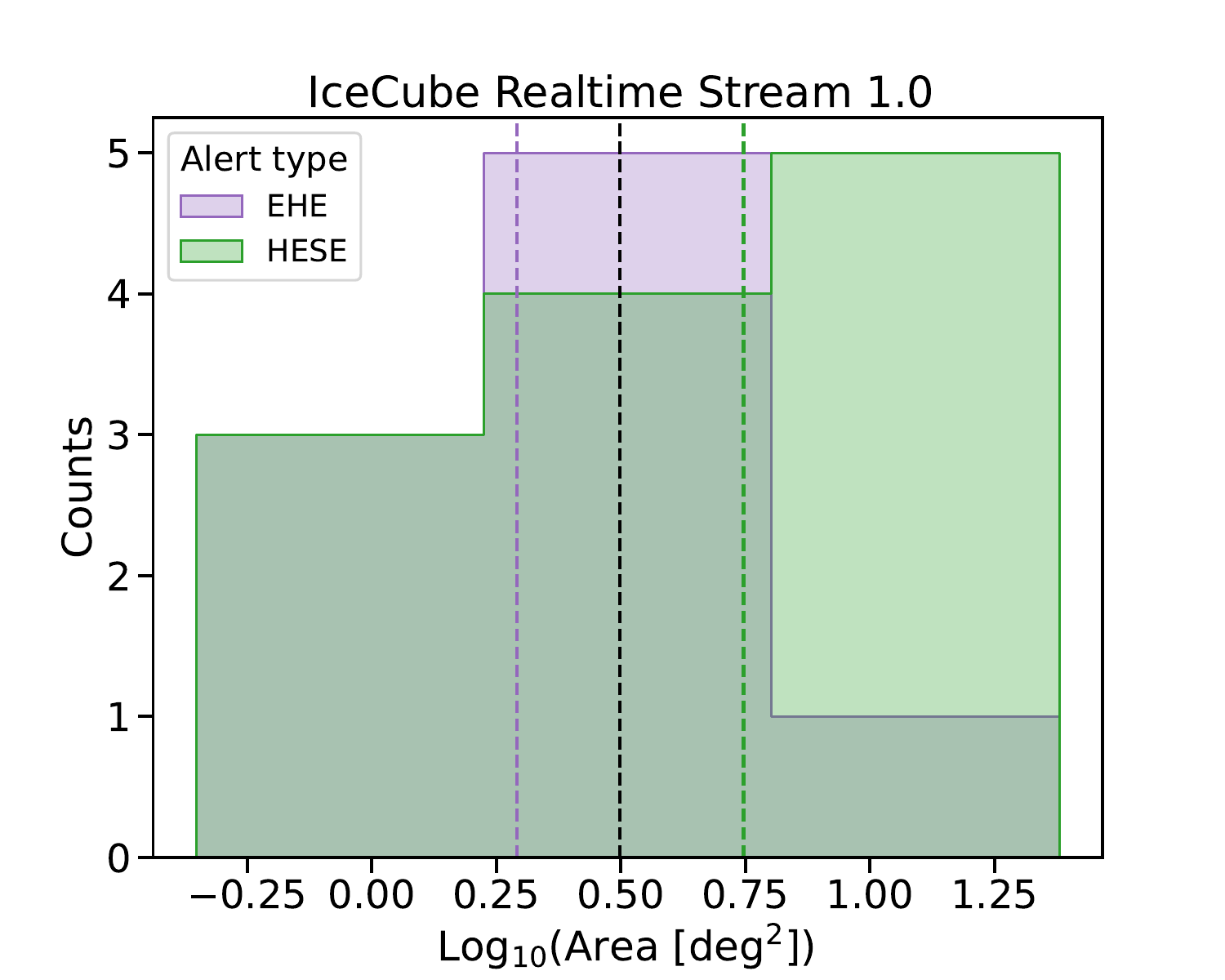}\includegraphics[width=0.5\linewidth]{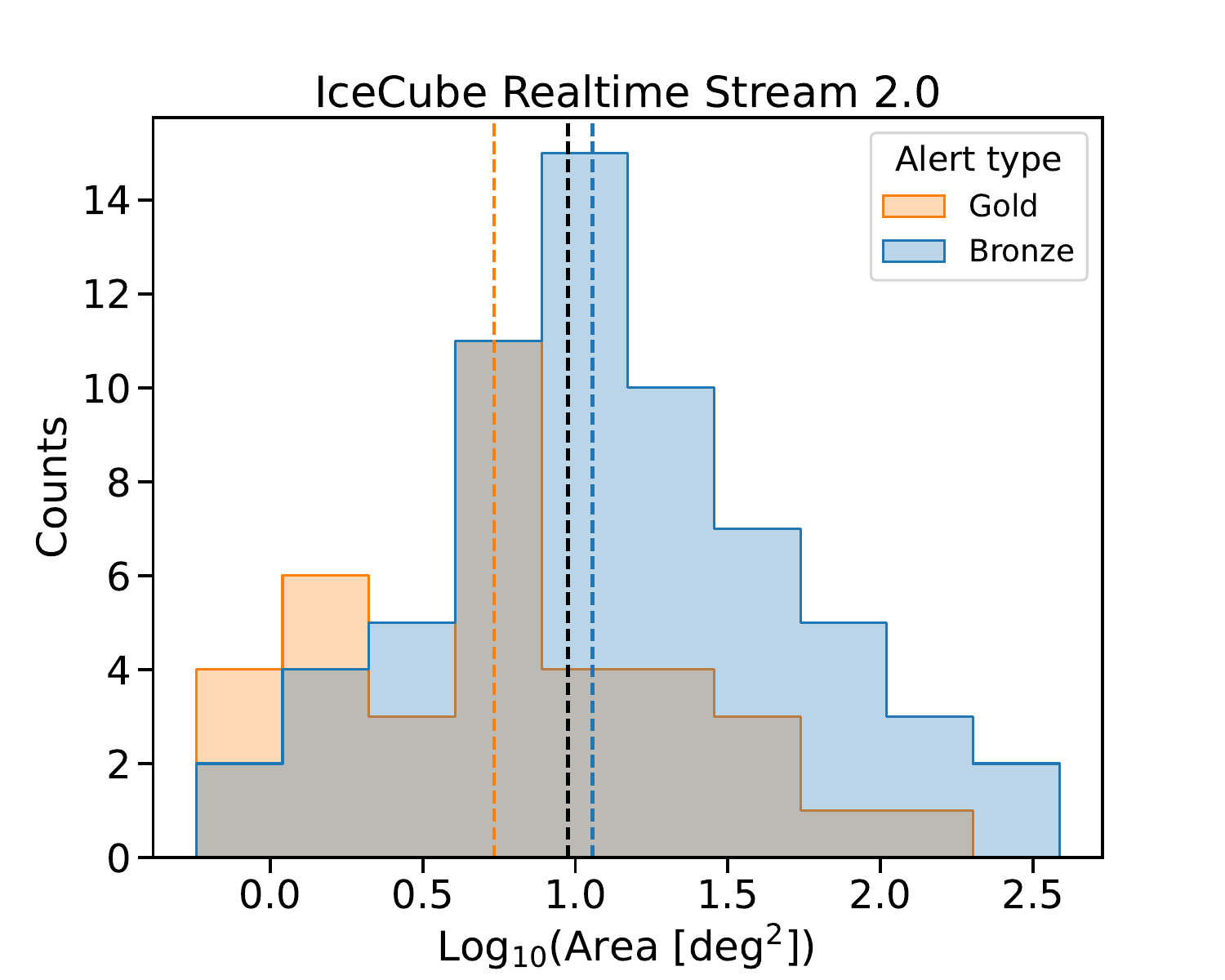}

    \caption{Distribution of 90$\%$ error region extensions for the IceCube Realtime Stream 1.0  \textit{(left)} and 2.0 \textit{(right)} colour-coded by alert type. The dashed colored lines in both plots show the median values for the each alert class, while the black dashed lines show the medians of the full samples of each alert stream.}
    \label{fig:errors_4fgl}
\end{figure*}

In June 2019, the IceCube Realtime Alert System was updated to its 2.0 version \citep{Blaufuss:2019fgv}. In this new stream, alerts are classified according to their average probability of being of astrophysical origin (signalness, \textit{s}). In the period considered for this work, between June 2019 and May 2023, the \textit{Fermi}-LAT observed all 101 alerts issued by IceCube. Among these, 37 were classified as Gold (\textit{s} > 50$\%$) and 64 as Bronze (\textit{s} > 30$\%$). The right panel of Figure \ref{fig:errors_4fgl} shows the distributions of the neutrino 90$\%$ containment regions for each alert classification, ranging from 0.57 deg$^{2}$ up to 385 deg$^{2}$. The median extension of the full sample is 9.5 deg$^{2}$ (black dashed line) while it is 5.4 deg$^{2}$ for the Gold sample and 11.4 deg$^{2}$ for the Bronze.

Of the 101 alerts, 44 are not coincident with any 4FGL-DR3 sources, while 17 are each coincident with a single 4FGL-DR3 candidate. The remaining 40 have several source candidates.\\

\subsection{Newly detected gamma-ray sources}
During realtime follow-up observations of high-energy neutrino alerts, new gamma-ray sources that do not already appear in the LAT catalogs are occasionally found and reported to the community. These sources are not necessarily causally connected to the neutrinos, and none of them were exhibiting increased gamma-ray activity at the time of the neutrino observations.

Nineteen new sources with a detection significance > 4$\sigma$ have been reported in the recent years, of which only 7 had a candidate associated counterpart. A general selection criterion for candidate counterparts is to have, within the 90$\%$ localization error of the LAT source, a blazar candidate detected in different catalogs. In Table \ref{tab:newsource_table}, we list these new sources along with the corresponding IceCube alert name and the tentative association. Only 4 of these sources have been reported in subsequent versions of the 4FGL-DR3 catalog (shown in boldface in Table \ref{tab:newsource_table}). Differences between the configuration of the follow-up analysis described in this work and that of the analysis of the catalog sources could explain why part of this sample is still not included in the \Fermi-LAT catalogs. Moreover, some of them could arise from systematic uncertainties on the models of the diffuse gamma-ray components that procude spurious residuals in the global fit of the ROI.  

\begin{table*}[h!]
\centering
\caption{New gamma-ray sources detected at > 4$\sigma$ that are in coincidence with IceCube alerts during realtime follow-up observations with \Fermi-LAT data. Sources in boldface are included in the latest version of the \Fermi-LAT 4FGL catalog used in this work.}
\begin{tabular}{llll}
\hline
\textbf{Alert Name} &     \textbf{\textit{Fermi}-LAT Name} &                      \textbf{Association} &   \textbf{ATel/GCN} \\
\hline
 IC190704A & \textbf{Fermi J1045.3+2751} &    1WHSP J104516.2+275133 & ATel\#12906 \citep{2019ATel12906....1G} \\
 IC191119A & \textbf{Fermi J1511.0+0550} &       NVSS J151100+054916 & ATel\#13306 \citep{2019ATel13306....1G} \\
 IC200109A & \textbf{Fermi J1055.8+1034} &                       - & ATel\#13402 \citep{2020ATel13402....1G} \\
 IC200523A & Fermi J2231.0+0034 & WISEA J223133.89+003312.8 & GCN\#27816 \citep{2020GCN.27816....1B} \\
 IC200530A & Fermi J1707.0+2528 &                       - &  GCN\#27879 \citep{2020GCN.27879....1B} \\
 IC200614A & \textbf{Fermi J0202.8+3132} &       NVSS J020242+313212 & ATel\#13811 \citep{2020ATel13811....1G} \\
 IC200911A & Fermi J0330.1+3743 &                       - & ATel\#14010 \citep{2020ATel14010....1G} \\
 IC200921A & Fermi J1256.9+2630 &                       - & ATel\#14038 \citep{2020ATel14038....1G} \\
 IC201021A & Fermi J1725.5+1312 &      1RXS 172314.4+142103 & ATel\#14111 \citep{2020ATel14111....1B} \\
 IC201114A & Fermi J0703.5+0505 &                       - & ATel\#14188 \citep{2020ATel14188....1G} \\
 IC210503A & Fermi J0931.9+3633 &  SDSS J093209.60+363002.6 & ATel\#14611 \citep{2021ATel14611....1G} \\
 IC210516A & Fermi J0610.5+0946 &                       - & ATel\#14639 \citep{2021ATel14639....1G} \\
 IC220115A & Fermi J2350.2+2620 &                       - & ATel\#15166 \citep{2022ATel15166....1G} \\
 IC220205A & Fermi J1420.7+1653 &                       - & ATel\#15211 \citep{2022ATel15211....1G} \\
 IC220624A & Fermi J1458.0+4119 &                       - & ATel\#15478 \citep{2022ATel15478....1G} \\
 IC220822A & Fermi J1810.1+2154 &                       - & ATel\#15570 \citep{2022ATel15570....1G} \\
 IC220918A & Fermi J0502.5+0037 &                       - & ATel\#15620 \citep{2022ATel15620....1G} \\
 IC230401A & Fermi J0030.2+0005 &           5BZB J0030-0000 & ATel\#15977 \citep{2023ATel15977....1G} \\
 IC230506A & Fermi J0320.2+2243 &                       - & GCN\#33745 \citep{2023GCN.33745....1G} \\
\hline
\end{tabular}
\label{tab:newsource_table}
\end{table*}

\subsection{Multiplet neutrino sources}\label{sec:multiplet}

As the number of neutrino alerts has increased and due to the larger average localizations of the current alert streams (see Figure \ref{fig:errors_4fgl}), there has been a steady increase in the number of 4FGL-DR3 sources that are coincident with the localizations of more than one event. Here, we consider the high-energy alerts observed during the nominal operations of the realtime streams, not imposing any constraints on the amount of time between consecutive alerts. This is different from the definition of neutrino multiplets adopted in other multiwavelength follow-up programs (e.g. \citealt{2017A&A...607A.115I}).
We find 14 sources coincident with the 90$\%$ localization regions of 2 neutrino alerts. Of these sources, 11 are associated with blazars, one is associated with the radio galaxy TXS 1516+064 and three are unassociated. All multiplet coincidences are listed in Table \ref{tab:multiplet_table}.
The last column of Tab. \ref{tab:multiplet_table} shows the extensions in deg$^{2}$ of the alerts coincident with the listed sources. Several sources are coincident with pairs of poorly reconstructed alerts, such as IC210608A and IC221210A. 

The most significant candidate blazar counterpart to a high-energy neutrino thus far, \TXS (also known as 4FGL J0509.4+0542), belongs to that list of multiplet sources. In contrast with the outstanding multiwavelength activity observed during the detection of IC170922A \citep{IceCube:2018dnn}, \TXS was not particularly active in gamma rays at the time of the detection of IC220918A \citep{2022ATel15620....1G}. Unlike the reconstructed direction of the EHE event IC170922A, the direction of the Bronze event IC220918A has a 90$\%$ containment of $\sim$51 deg$^{2}$, leaving plenty of room for a random coincidence.\\

\begin{table*}[h!]
\centering
\caption{Sources that are spatially coincident with multiple high-energy neutrino alerts. The listed multiwavelength associations and source classes are those reported in the 4LAC-DR3 catalog.}
\begin{tabular}{llrll}
\hline
        \textbf{4FGL Name} &                  \textbf{Association} & \textbf{Class} &                \textbf{IceCube names} & \textbf{Area (deg$^{2}$)} \\
\hline
J2306.6+0940 &                            - &     - &        IC190619A, IC220424A & 28, 5\\
J2308.9+1111 &             MG1 J230850+1112 &    bcu &       IC190619A, IC230201A & 28, 18\\
J1518.6+0614 &                 TXS 1516+064 &   rdg &       IC191119A, IC200410A & 60, 380\\
J1523.2+0533 &          NVSS J152312+053357 &   bll &       IC191119A, IC200410A & 60, 380\\
J0258.1+2030 &             MG3 J025805+2029 &   bll &       IC191231A, IC211125A & 38, 24\\
J1747.6+0324 &                            - &     - &       IC210510A, IC220205B & 4, 1\\
J2149.7+1917 &                 TXS 2147+191 &   bcu &       IC210608A, IC221210A & 110, 250\\
J2200.1+2138 &                 TXS 2157+213 &   bll &       IC210608A, IC221210A & 110, 250\\
J2218.6+1941 & GALEXASC J221854.64+193841.6 &   bcu &       IC210608A, IC221210A & 110, 250\\
J2225.6+2120 &                  PKS 2223+21 &  fsrq &       IC210608A, IC221210A & 110, 250\\
J2236.9+1839 &                            - &     - &       IC210608A, IC221210A & 110, 250\\
J2243.9+2021 &                RGB J2243+203 &   bll &       IC210608A, IC221210A & 110, 250\\
J2248.9+2107 &                 PKS 2246+208 &  fsrq &       IC210608A,  IC221210A & 110, 250\\
J0509.4+0542 &                TXS 0506+056 &   bll &      IC170922A, IC220918A & 1, 51\\
\hline
\end{tabular}

\label{tab:multiplet_table}
\end{table*}

\subsection{Multiplets in archival and realtime alerts}
\label{sec:multiplet_wICECAT}
Prior to the introduction of its Realtime Alert System, IceCube observed 35 additional high-energy neutrinos (11 HESE, 24 EHE) that comprise an archival list
 \citep{IC_Archival_Alerts}. When we include those archival events in our analysis, we find only one additional double coincidence. This is the BL Lac object GB6 J1040+0617, which was already identified as the counterpart of the HESE neutrino IC141209A \citep{2019ApJ...880..103G}. This source was not observed to be flaring at the time of the detection of the second coincident neutrino event, the Bronze event IC220627A. Additionally, no other sources were significantly detected in the region within short timescales \citep{2022GCN.32301....1G}.

In addition to the list of archival HESE/EHE alerts, we consider also the ICECAT-1 catalog of IceCube track alerts \citep{2023ApJS..269...25A}. This catalog contains realtime and archival events from 2011 to 2020 that would have been selected as Gold or Bronze high-energy alerts. From the sample of 275 alerts listed in this catalog, 225 are not already listed in the samples considered in Sections. \ref{sec:fup_res1} and \ref{sec:fup_res2}. In including those neutrino events, the number of 4FGL-DR3 sources coincident with multiple neutrino alerts increases to 115, and sources with up to 4 spatial coincidences appear. In Table \ref{tab:multiplet_table_wIceCAT}, the 11 sources that are spatially coincident with more than 2 high-energy neutrinos are listed. As already seen in the previous samples, the majority of these multiplet coincidences are caused by neutrino events with large uncertainties in their reconstruction, with areas in this sample that reach several hundreds of square degrees.\\
One source is coincident with four different neutrino alerts. This is 4FGL J2226.8+0051, associated with the FSRQ PKS B2224+006 (also known as 4C +00.81) at a redshift of $z = 2.25$ \citep{2017ApJS..233...25A}. As listed in Table \ref{tab:multiplet_table_wIceCAT}, this multiplet coincidence shares three neutrino alerts with three additional sources. The sizes of the 90$\%$ error contours of these alerts range from 2.7 deg$^{2}$ up to 90.6 deg$^{2}$; therefore each of these coincidences has a high chance of being a random coincidence. 

We note that the candidate neutrino-emitting blazar \TXS is correlated with an additional event from the ICECAT-1 sample, the Bronze archival event IC190317A. According to the light curve of the source that is included in the \Fermi-LAT Light Curve Repository\footnote{\url{https://fermi.gsfc.nasa.gov/ssc/data/access/lat/LightCurveRepository/source.html?source_name=4FGL_J0509.4+0542}} \citep{2023ApJS..265...31A}, IC190317A is detected after the end of a broader period of gamma-ray flaring activity that began in 2017. During this period, the average flux of the source was comparable to its 4FGL value.

\subsection{Estimating expected random coincidences}
Given the observed coincidences between 4FGL-DR3 sources and multiple neutrino alerts, it is essential to estimate the expected rate of such coincidences to occur by chance. To do so, we simulate 10$^{4}$ samples of neutrino alerts starting from the catalogs described above and by scrambling their right ascensions while keeping their declinations and 90$\%$ error contours fixed. This preserves the distribution of alerts as a function of declination which reflects the sensitivity of IceCube. By repeating the search for coincident 4FGL-DR3 sources for each of the simulated catalogs, we can determine the probability for a source to be coincident with one or multiple neutrino alerts.

Starting from the sample of 122 alerts issued in the realtime programs (see Sections \ref{sec:fup_res1} and \ref{sec:fup_res2}), we find that the probability for a 4FGL-DR3 source to be randomly coincident with two or more alerts is $p$($N \geq$ 2) = 3$\times$10$^{-2}$. This gives a total expected number of 21 random coincidences. The 14 4FGL-DR3 sources coincident with two neutrino alerts and listed in Table \ref{tab:multiplet_table} are therefore consistent with the expected number of random coincidences.
When we repeat this simulation for the full sample of 382 alerts which includes the archival events described in Section \ref{sec:multiplet_wICECAT}, we find that the expected number of sources randomly coincident with 3 (4) or more alerts is 18 (2). Therefore, the 11 observed triple-coincidences and the coincidence of 4FGL J2226.8+0051 with 4 neutrino alerts (Table \ref{tab:multiplet_table_wIceCAT}) are also consistent with random chance.

\begin{table*}[h!]
\centering
\caption{Sources that are spatially coincident with multiple high-energy neutrino alerts from the expanded sample that includes events from the ICECAT-1 catalog.}

\begin{tabular}{llll}
\hline
        \textbf{4FGL Name} &                  \textbf{Class} &                \textbf{IceCube names} & \textbf{Area (deg$^{2}$)} \\
\hline
J0203.9+8120 &             - &            IC190629A, IC140103A, IC140410A & 161.6, 837.6, 3143.8 \\
J0228.1+8208 &           bcu &            IC190629A, IC140103A, IC140410A & 161.6, 837.6, 3143.8 \\
J2223.3+0102 &    bll &            IC200523A, IC110807A, IC180612A &      90.6, 5.7, 70.4 \\
J2226.6+0210 &     bcu &            IC200523A, IC110807A, IC180612A &      90.6, 5.7, 70.4 \\
J2226.8+0051 &             fsrq & IC200523A, IC110807A, IC140114A, IC180612A & 90.6, 5.7, 2.7, 70.4 \\
J2227.9+0036 &              bll &            IC200523A, IC140114A, IC180612A &      90.6, 2.7, 70.4 \\
J0506.9+0323 &         bcu &            IC220918A, IC161117A, IC190317A &     50.9, 13.9, 95.0 \\
J0509.4+0542 &                bll &            IC220918A, IC170922A, IC190317A &      50.9, 1.3, 95.0 \\
J1019.7+0511 &         bcu &            IC130627A, IC170308A, IC190415A &       4.9, 6.0, 33.8 \\
J0839.7+3540 &         bll &            IC140122A, IC140223A, IC181121A &   55.1, 368.2, 111.2 \\
J0244.7+1316 &              bcu &            IC161103A, IC170824A, IC180613A &      3.2, 18.2, 91.5 \\
\hline
\end{tabular}

\label{tab:multiplet_table_wIceCAT}
\end{table*}

\section{The case of IceCube-211208A and PKS 0735+178}
\label{sec:pks0735}
In this section, we present the results of a particular neutrino blazar candidate that was barely outside of the localization contour of an IceCube alert, but is considered interesting because of its flaring activity and the spatial coincidence with neutrinos from other observatories within a 4-day time frame.
On 8 December, 2021 the Bronze neutrino alert IC211208A was issued with a reconstructed direction of RA = 114.52 (+2.82, -2.50) deg, Dec. = 15.56 (+1.81, -1.39) deg (90$\%$ containment) and an estimated energy of 171 TeV \citep{2021GCN.31191....1I}. The signalness for this event was estimated to be $\sim$50$\%$, and two LAT sources were found within its 90$\%$ uncertainty region. From the \Fermi-LAT observations of the sky region around this event, we found that neither of these two sources were significantly detected within timescales of 1-month and 1-day prior to the neutrino arrival. However, the analysis of the region showed that a gamma-ray source, located 20 arcmin outside of the IceCube 90$\%$ uncertainty countour, was significantly detected (> 5$\sigma$) within short timescales. This was the source 4FGL J0738.1+1742, associated with the BL Lac object PKS 0735+178 at $z = 0.45$ \citep{2012A&A...547A...1N}. We detected the source with an average flux of $F = (3 \pm 1)\times 10^{-7}$ ph cm$^{-2}$ s$^{-1}$ ($\sim 6$ times the average 4FGL value), integrated over the 24 hours before the neutrino detection \citep{2021ATel15099....1G}. \\
Interestingly, on 14 December 2021, the Baikal-GVD collaboration reported the observation of a cascade-like event with reconstructed direction of RA = 119.44 deg, Dec = 18.00 deg (5.5 deg radius, 50$\%$ containment) and an estimated energy of 43 TeV. The event was detected 3.95 hours after the IceCube event and is consistent with the position of the blazar PKS 0735+178 and that of IC211208A. The Baikal-GVD collaboration estimated that the chance coincidence between the neutrino and the source can be excluded at the 2.85 $\sigma$ level \citep{2021ATel15112....1D}.\\
A third neutrino event coincident with PKS 0735+178 and the other two high-energy neutrinos was detected on 4 December 2021 by the Baksan Underground Scintillation Telescope (BUST), 4 days before the IceCube event. The reported neutrino arrival direction was RA = 116.5 deg, Dec = 16.6 deg (2.5 deg, 50\% PSF containment). The energy was estimated to E $>$ 1 GeV. \citet{2021ATel15143....1P} report that the chance coincidence between the neutrino and the source can be excluded with a significance of 3$\sigma$.

\begin{figure}[h!]
    \centering
	\includegraphics[width=\linewidth]{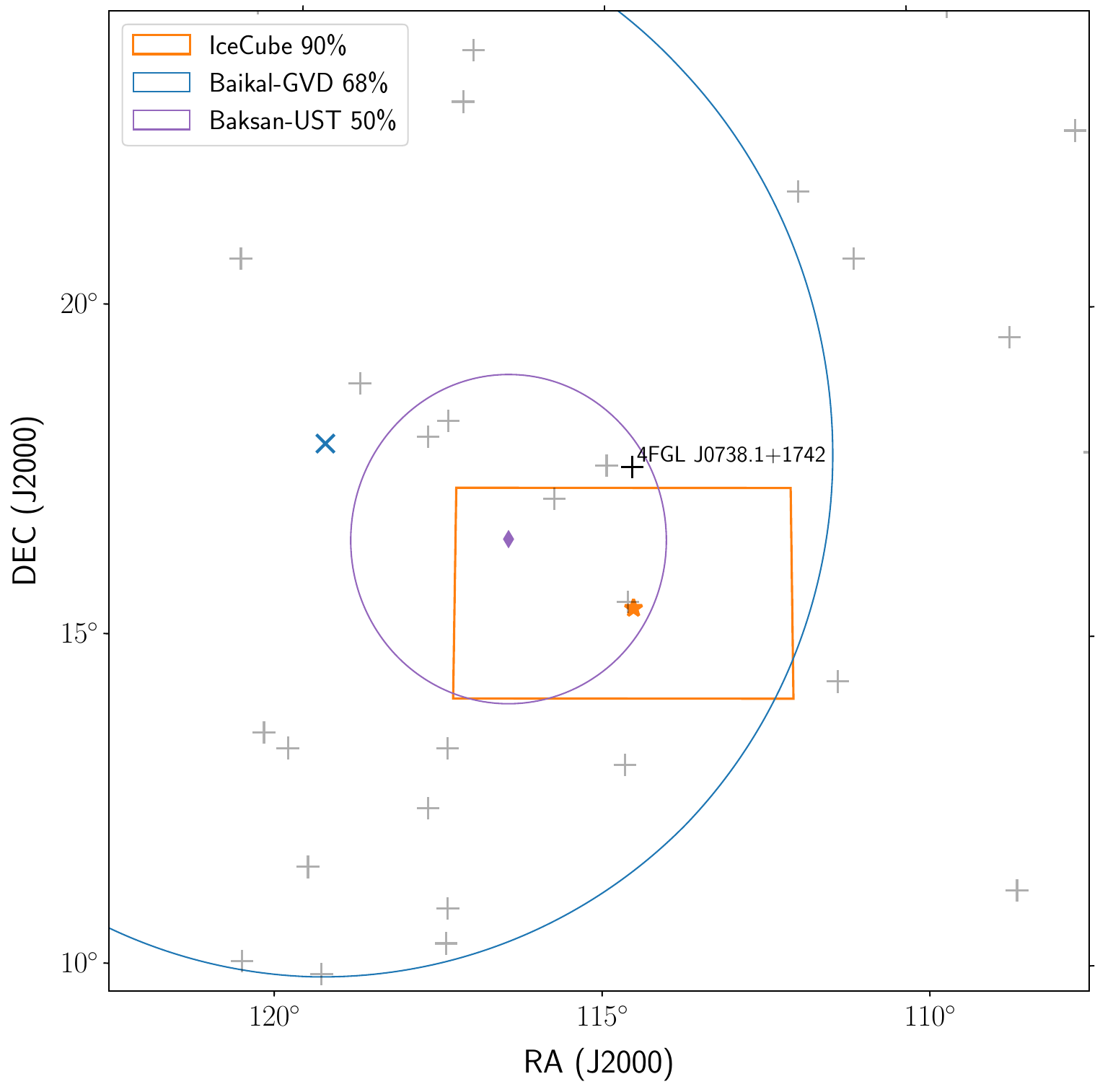}\\
    
	\caption{Sky map centered on the position of PKS 0735+178 (4FGL J0738.1+1742). The patches show the 90$\%$ IceCube localization (orange), the 68$\%$ Baikal-GVD localization (blue) and the 50$\%$ Baksan-UST localization (purple), respectively. They grey markers are all 4FGL sources in the region.}
	\label{fig:pks0735_Map}
\end{figure}

The sky region of PKS 0735+178 with the localizations of the three neutrinos is shown in Fig. \ref{fig:pks0735_Map}. Recent studies have demonstrated that the current correction values adopted by IceCube to determine the 90$\%$ error contours of realtime alerts do not fully account for systematic effects in the majority of cases (e.g. \citealt{2021arXiv210708670L}). Future reprocessing of the data may lead to shifts in the arrival direction of this particular event such that its localization contour may include PKS 0735+178, as well as the two neutrino events observed by the other detectors. In such a case, PKS 0735+178 would have a triple neutrino coincidence in both space and in time.

\begin{figure*}
\sidecaption    
    \includegraphics[width=12cm]{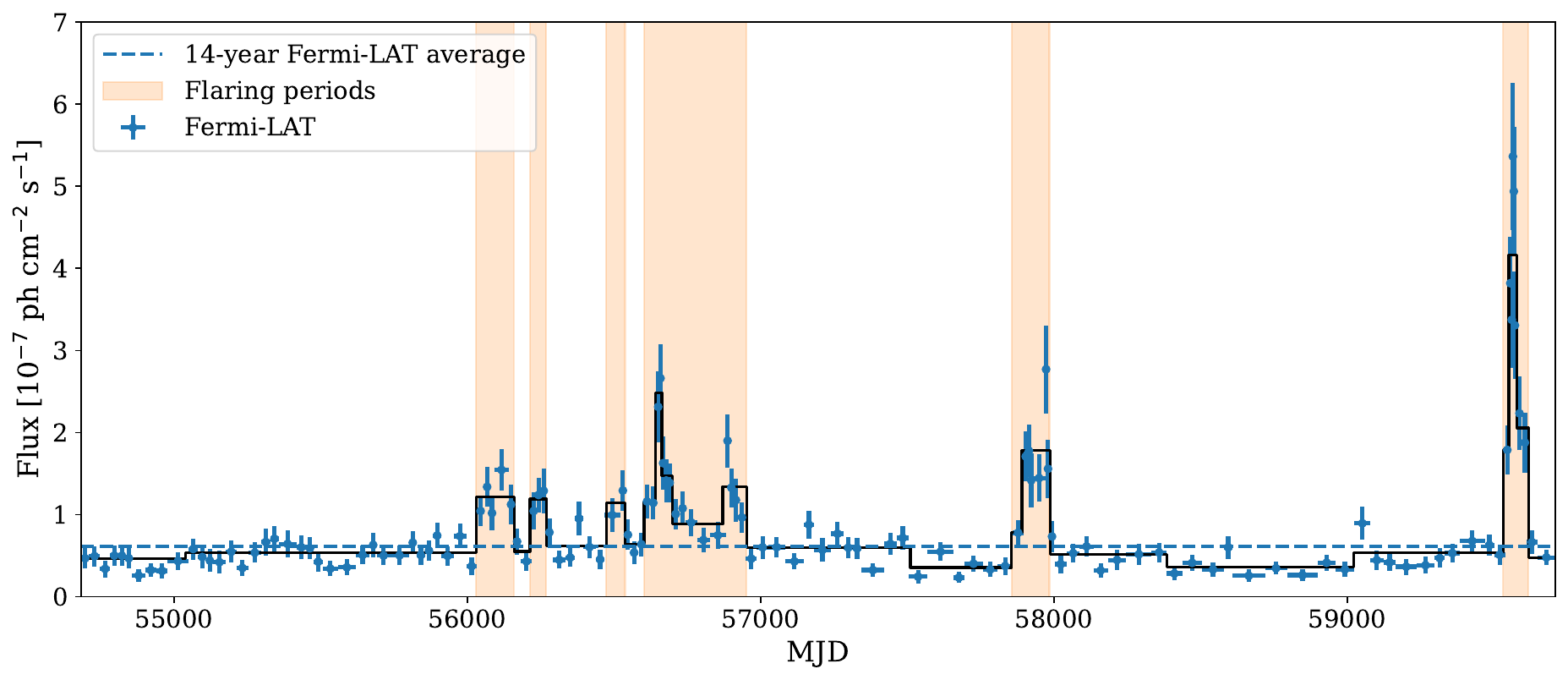}
        \caption{\Fermi-LAT adaptively binned light curve of PKS 0735+178 from August 2008 to May 2022. The shaded orange periods indicate the time intervals in which the source is observed with an integrated gamma-ray flux that is above the 14-year average in the Bayesian blocks representation. }
	\label{fig:pks0735LC_fullmission}
\end{figure*}

\begin{figure*}
\sidecaption
    \includegraphics[width=12cm]{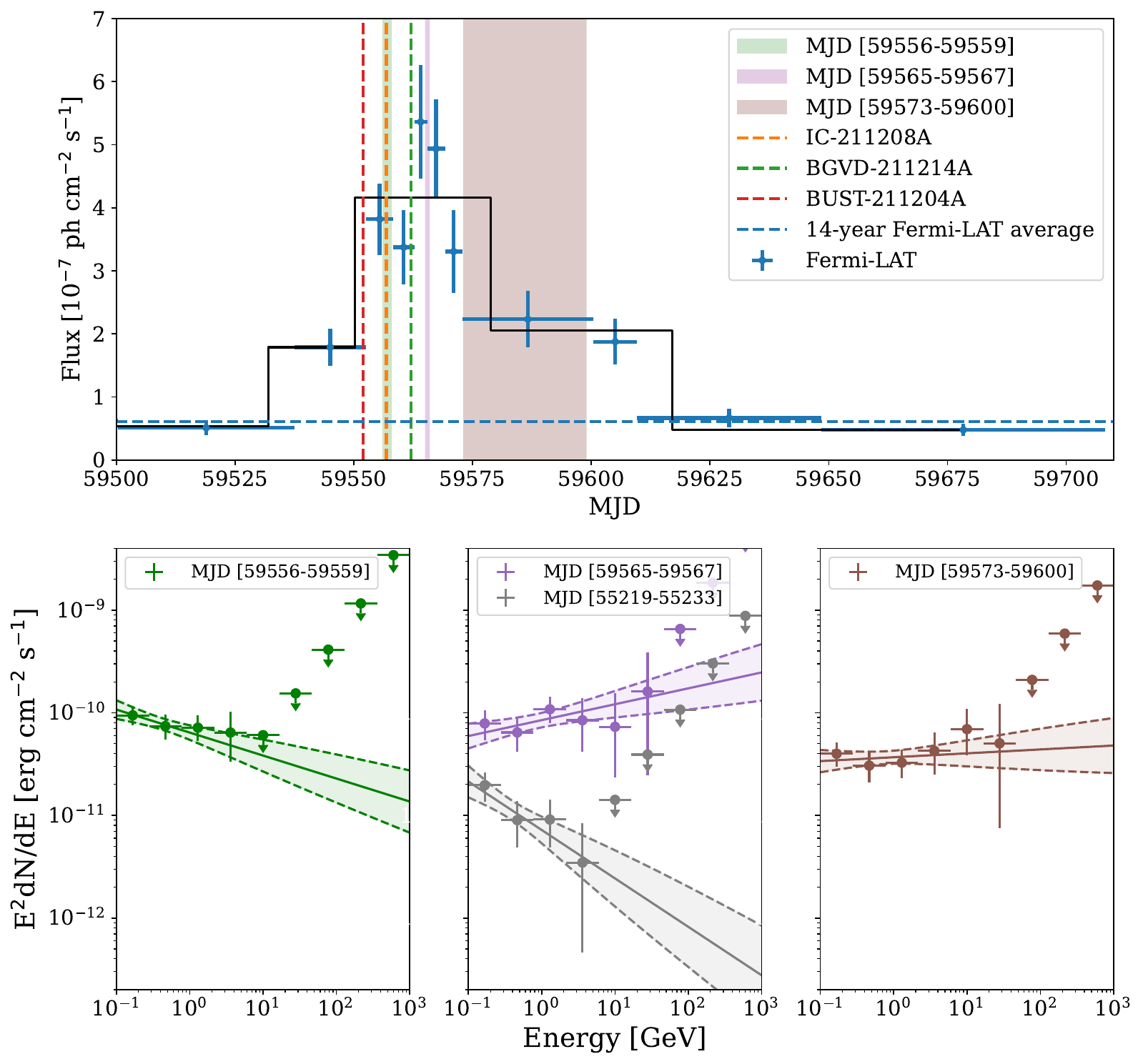}
        \caption{\Fermi-LAT adaptively binned light curve of PKS 0735+178 from October 2021 to May 2022. The upper panel shows the integrated fluxes (blue markers) and the dashed vertical lines indicate the arrival times of the neutrinos, while the shaded regions indicate the periods for which the SEDs shown in the lower panel are calculated. The gray SED in the middle lower panel corresponds to the quiescent state between MJD 55219-55233. }
	\label{fig:pks0735LC}
\end{figure*}

The gamma-ray flare of PKS 0735+178 that occurred at the time of the detection of IC211208A was the brightest activity from this source observed by the \Fermi-LAT up to that point, as shown in the 14-year light curve between August 2008 and May 2022 (Fig. \ref{fig:pks0735LC_fullmission}). An analysis of the light curve using Bayesian Blocks \citep{2013ApJ...764..167S} reveals only 5 more periods of enhanced activity above the 14-year average (shaded orange areas). With the flare that coincided with the detections of the three neutrino events, the active periods in which this source exhibited fluxes above the average constitute a duty cycle of $\sim$16\%. The source was then in a quiescent state between March and July 2022, followed by several months of renewed gamma-ray activity in the past year, as shown in the continously updated data from the \Fermi Light Curve Repository.\\ 
The upper panel of Figure \ref{fig:pks0735LC} shows the adaptively binned gamma-ray light curve (following the method of \citealt{2012A&A...544A...6L}) of \Fermi-LAT data from 13 October 2021 to 9 May 2022 (MJD 59500 - 59708). In the lower panels of the figure, we show the \Fermi-LAT spectral energy distributions (SED) of three selected periods with good multiwavelength coverage: simultaneous with the IceCube neutrino arrival (green), around the peak of the gamma-ray flare (purple), and in the decay phase of the flare (brown). For comparison, we also show a historical quiescent state (gray) from 23 January 2010 to 6 February 2010. The gamma-ray SEDs in these short intervals, selected from the recent period of gamma-ray activity, are well-described using a power-law model with measured spectral indices of $\Gamma_{1}$ = 2.22$\pm$0.13 (MJD 59556-59559, green), $\Gamma_{2}$ = 1.84$\pm$0.12 (MJD 59565-59567, purple) and $\Gamma_{3}$ = 1.96$\pm$0.12 (MJD 59573-59600, brown). The spectral indices $\Gamma_{2}$ and $\Gamma_{3}$ are significantly harder than the one observed in the quiescent state during MJD 55219-55233 (gray) of $\Gamma_{4}$ = 2.47$\pm$0.26.
Recent theoretical models of the broadband emission of PKS 0735+178 observed during this flaring period predicted neutrino fluxes that are consistent with the observations~\citep{2023MNRAS.519.1396S,Omeliukh:2023Ic,2023ApJ...954...70A}.

\section{Testing the gamma-neutrino connection in blazars}
\label{sec:gamma_neutrino_connection}
\subsection{Method}
Although the number of coincidences between multiple neutrino alerts and gamma-ray sources is statistically consistent with the random background hypothesis, a correlation between observed neutrinos and potential candidates might still be found when considering a correlation between the gamma-ray and the neutrino flux. The correlation hypothesis that is tested in this section does not require an increase in gamma-ray activity from the source at the time of the neutrino detection (as suggested by the exceptional case of \TXS). Rather, the hypothesis only considers the average gamma-ray brightness of these objects. Such a correlation could point to a sub-population of candidates with peculiar environments and emission region characteristics that are conducive to neutrino production (i.e. dense targets of dust, gas and photon fields). Such a scenario has been extensively studied in the case of the coincidence between \PKS and IC190730A, where the source was observed in a quiescent state at the time of the neutrino detection but can still be considered a promising neutrino emitter because of its intrinsic properties \citep{2021ApJ...912...54R,2021JCAP...10..082O}. 

In order to perform this study, it is crucial to select a sample of candidates where the probability of chance coincidence is reduced as much as possibile. These random coincidences arise mostly from poorly reconstructed neutrino arrival directions and from incorrectly classified atmospheric neutrinos. The former can be reduced by selecting a subsample of well-reconstructed neutrinos, while we will consider the latter an irreducible contamination given the moderate signalness of the events selected by the realtime program.

In the following, we use an approach similar to the one adopted in \cite{2019ApJ...880..103G} and \cite{2020ApJ...893..162F} in order to select coincidences from a sample of relatively well-reconstructed neutrino alerts. For the threshold for the extension of the error regions, we consider the median angular uncertainty of 5.4 deg$^{2}$ observed in the sample of Gold alerts for alerts (realtime and archival) classified as Gold/Bronze. For alerts (realtime and archival) classified as HESE/EHE, we use the median angular uncertainty of the HESE sample of 5.6 deg$^{2}$. For this selection, we define well-reconstructed alerts as all events with extensions below the aforementioned thresholds.

Since gamma-ray blazars dominate the observed coincidences with single high-energy neutrinos, and since they are prime candidate neutrino sources, we consider sources from the Fourth \Fermi-LAT Catalog of Active Galactic Nuclei (4LAC-DR2, \citealt{2020ApJ...892..105A}, \citealt{2020arXiv201008406L}). We apply the selection cut on the 90$\%$ error contours, and we select only alerts coincident with a single 4LAC source. After the selection, the sample includes 3 alerts from the HESE/EHE realtime stream, 4 from the Gold/Bronze realtime stream and 8 from the archival samples (3 from archival HESE/EHE alerts and 5 from archival alerts in ICECAT-1). The 15 4LAC sources that are the sole candidate counterpart to each of the relatively well-reconstructed neutrino alerts with a single 4LAC coincidence are listed in Table \ref{tab:coincidence_table}. 

\begin{figure*}
\sidecaption
    \includegraphics[width=12cm]{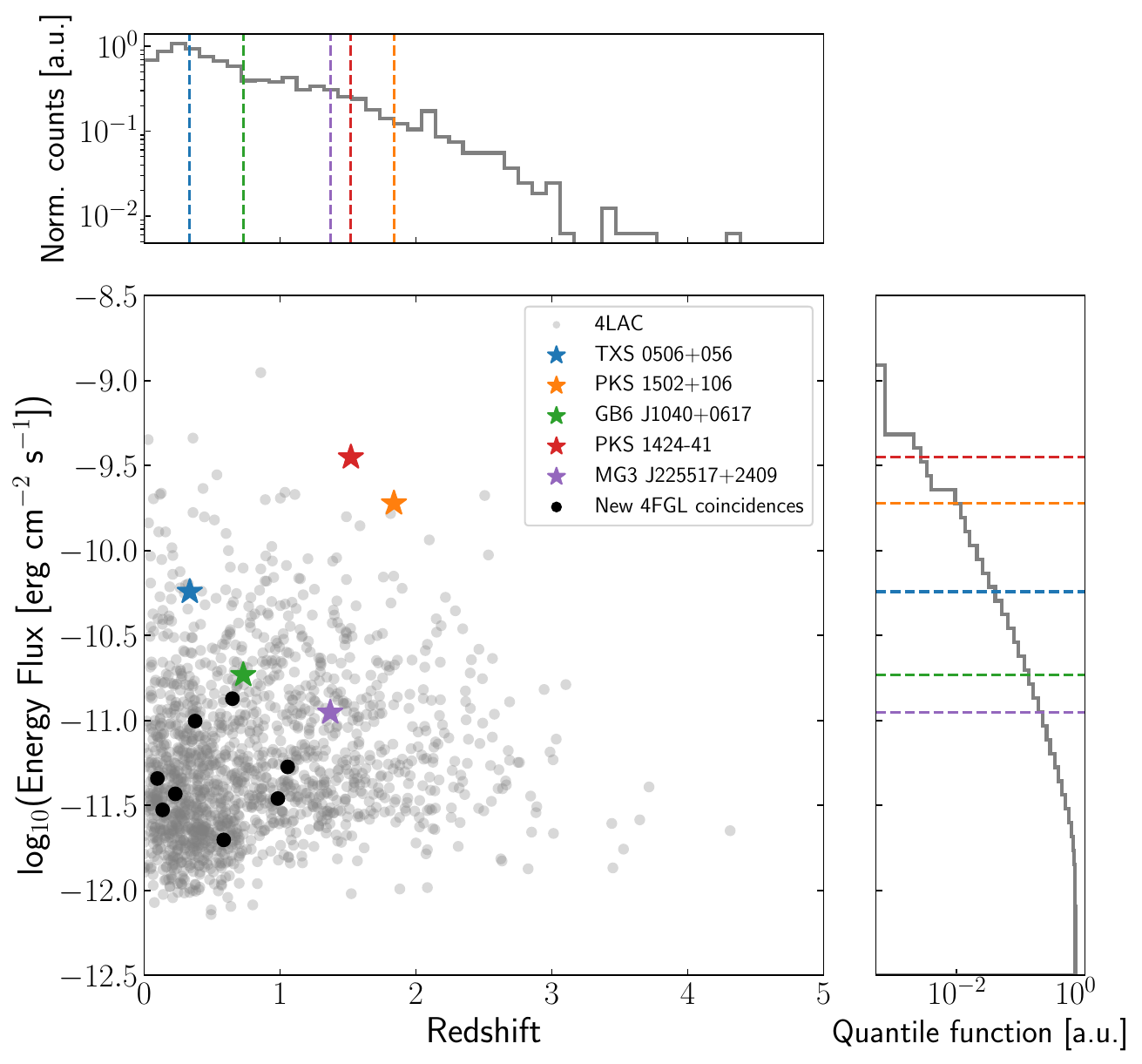}
        \caption{Comparison of candidate neutrino blazars with all blazars in the 4LAC AGN sample adapted from \cite{2020ApJ...893..162F}. Star-shaped markers indicate candidate neutrino blazars already studied in the literature, and black circles show the new coincident sources (with identified counterparts) from Table \ref{tab:coincidence_table}. \TXS and \GB are also listed in Table \ref{tab:coincidence_table} as their coincident neutrinos satisfy the criteria of well-reconstructed alerts.}
        \label{fig:EfluxRedshift}
\end{figure*}

Figure \ref{fig:EfluxRedshift} shows the 2D distribution of gamma-ray energy fluxes (in the energy range from 100 MeV to 100 GeV) and redshift for all the blazars in 4LAC as grey dots (see also \citealt{2020ApJ...893..162F}). We show the 4LAC sources that are coincident with well-reconstructed neutrino alerts and have measured redshift as black dots, and we use colored stars for the sources that have been suggested in the literature as neutrino blazars on the basis of their multiwavelength properties in addition to their coincidence with high-energy neutrinos. Additional neutrino blazar candidates were omitted from Figure \ref{fig:EfluxRedshift} because they lack measured redshifts. In the side plots, we show the inverse cumulative distribution function of the gamma-ray energy fluxes (right) and the normalized counts distribution of the redshifts (top). 

In addition to the selection described above, we include (and show in Figure \ref{fig:EfluxRedshift}) two additional neutrino blazar candidates motivated by strong multiwavelength coincidences found among neutrino detections published in various archival samples of IceCube data. The blazars are the FRSQ PKS 1424-41, coincident with the HESE cascade event IceCube-35 \citep{2016NatPh..12..807K} and the BL Lac MG3 J225517+2409, coincident with the event IC100608A and a hot spot from a search conducted by ANTARES \citep{2011NIMPA.656...11A} for point sources consistent with the directions of \Fermi-LAT blazars \citep{2021ApJ...911...48A}. Finally, we include the FSRQ PKS 1502+106 coincident with the gold alert IC190730A. IC-190730A has a 90$\%$ uncertainty area of 5.5 deg$^{2}$, slightly larger than the observed median for gold alerts of 5.4 deg$^{2}$. Nevertheless, PKS 1502+106 has been proven to be a valid candidate for IC190730A by several works (e.g. \citealt{2021ApJ...912...54R}, \citealt{2021JCAP...10..082O}). A total of 18 sources are in the final sample of neutrino blazar candidates.

\citet{2020ApJ...893..162F} presented a correlation study between gamma-ray blazars and high-energy neutrinos. The study used the distribution of gamma-ray energy fluxes of sources that were coincident with well-reconstructed neutrino alerts to probe the more general relation,

\begin{equation}\label{eq:lnu_lgamma}
 L_{\nu} \propto L_{\gamma}^{\alpha},
\end{equation}
between the correspondent average neutrino and gamma-ray luminosities ($L_{\nu}$ and $L_{\gamma}$, respectively). Although from single neutrino coincidences it is not possible to calculate $L_{\nu}$, this relation can be tested by checking how compatible the $L_{\gamma}$ distribution of the candidate neutrino blazars is with the expected distribution of gamma-ray blazars for each value of $\alpha$. \\
\citet{2020ApJ...893..162F} showed mild ($p$ = 64\%) evidence that neutrino-emitting blazar candidates are statistically compatible with the hypothesis of a linear correlation between the neutrino and gamma-ray energy fluxes. Here, we repeat this study over a larger sample of coincidences and with two main variations:

\begin{itemize}
    \item We consider a grid of values for the correlation power-law index, $\alpha$, from 0 (signifying a non-correlation) to 2 (signifying a quadratical correlation), in steps of 0.125. 
    \item We also perform the test on the distribution of physical luminosities for the subsample of sources with measured redshifts.
\end{itemize}

In the first test (\textit{T1}), we use the gamma-ray energy flux as a proxy for the luminosity and use the full sample of 18 sources. Following the optical classification of blazars, the sample is divided into 9 BL Lacs, 3 FSRQs, and 6 blazars of uncertain type (BCU). This sample is tested against the full sample of 3137 blazars (1131 BL Lacs, 694 FSRQs, and 1312 BCU) in the 4LAC-DR2 catalog. We use a Kolmogorov-Smirnov (\textit{K-S}) test \citep{1958ArM.....3..469H} to compare the two samples and statistically quantify how likely they are to belong to the same population. For each value of the index $\alpha$, 10$^{4}$ flux values (F) are sampled from the full 4LAC distribution with a sampling probability defined as

\begin{equation}\label{eq:norm_p_factor}
 p_{i} = \frac{F_{i}^{\alpha}}{\sum_{i}{F^{\alpha}_{i}}}
\end{equation}
to create a distribution to be tested against the neutrino blazar candidates by a two-sample \textit{K-S} test.

In the second test (\textit{T2}), we use the distribution of gamma-ray luminosities and consider only the BL Lac and FSRQ samples. Hence, the neutrino blazar sample is reduced to 9 objects (6 BL Lacs and 3 FSRQs) with measured redshifts, and the 4LAC sample is also reduced to 697 BL Lacs and 691 FSRQs. Each test is repeated 10$^{3}$ times for each value of $\alpha$, and the average and standard deviation of all the p-values from the \textit{K-S} tests is considered. As an additional test to ensure the robustness of our \textit{K-S} tests in both $T1$ and $T2$, we repeat the same exercise 10$^{3}$ times for each value of $\alpha$ using randomly sampled distributions of dummy neutrino blazars with the same size as the original samples.

\begin{table*}[h!]
\centering
\caption{Selection of 4LAC-DR2 sources that are coincident with well-reconstructed realtime alerts from Section \ref{sec:gamma_neutrino_connection}. Two sources are highlighted in boldface, 4FGL J0509.4+0542 (TXS 0506+056) and 4FGL J1040.5+0617 (GB6 J1040+0617) that are already known neutrino blazar candidates.}
\begin{tabular}{lllllll}
\hline
\textbf{4FGL Name} & \textbf{Class}\footnotemark[1]         & \textbf{E.Flux [erg cm$^{-2}$ s$^{-1}$]}\footnotemark[2] & \textbf{Redshift}     & \textbf{Event} & \textbf{Type} & \textbf{Sig.} \\ \hline
\textbf{J0509.4+0542}  & BL Lac                      & (5.72 $\pm$ 0.20)$\times$10$^{-11}$      & 0.3365                  & IC170922A      & EHE                 & 0.51                \\
 \textbf{J1040.5+0617}  & BL Lac                    & (1.85 $\pm$ 0.08)$\times$10$^{-11}$      & 0.73                & IC141209A      & HESE               & \multicolumn{1}{c}{-}                \\
J0658.6+0636  & BCU                       & (3.70 $\pm$ 0.73)$\times$10$^{-12}$      &0.23\footnotemark[3]  & IC201114A      & Gold                 & 0.56                \\
J1342.7+0505  & BL Lac                    & (2.98 $\pm$ 0.49)$\times$10$^{-12}$      & 0.13663               & IC210210A      & Gold                 & 0.65                \\
J0127.2+0324    & BL Lac  & (4.63 $\pm$ 0.47)$\times$10$^{-12}$      & \multicolumn{1}{c}{-} & IC220202A      & Gold               & 0.21 \\
J2322.7+3436    & BL Lac  & (4.90 $\pm$ 0.45)$\times$10$^{-12}$      & 0.098 & IC221223A      & Gold               & 0.79 \\
J0244.7+1316    & BCU  & (1.93 $\pm$ 0.67)$\times$10$^{-12}$      & 0.9846\footnotemark[4] & IC161103A      & Bronze               & 0.31 \\
J1758.7-1621    & BCU  & (8.04 $\pm$ 1.41)$\times$10$^{-12}$      & \multicolumn{1}{c}{-} & IC190221A      & HESE               & 0.37 \\
J1744.2-0353    & FSRQ  & (5.94 $\pm$ 1.19)$\times$10$^{-12}$      & 1.057 & IC110930A      & EHE               & \multicolumn{1}{c}{-} \\
J1258.7-0452    & BL Lac  & (2.07 $\pm$ 0.43)$\times$10$^{-12}$      & 0.586 & IC150926A      & EHE               & \multicolumn{1}{c}{-} \\
J1008.0+0620    & BL Lac  & (9.58 $\pm$ 0.71)$\times$10$^{-12}$      & 0.65 & IC190221A      & Bronze               & 0.40 \\
J1506.6+0813    & BL Lac  & (9.36 $\pm$ 0.76)$\times$10$^{-12}$      & 0.376 & IC121115A      & Bronze               & 0.32 \\
J0604.9-0000    & BCU  & (4.79 $\pm$ 0.84)$\times$10$^{-12}$      & \multicolumn{1}{c}{-} & IC130822A      & Bronze               & 0.30 \\
J0436.2-0038    & BCU  & (1.06 $\pm$ 0.28)$\times$10$^{-12}$      & \multicolumn{1}{c}{-} & IC180608A      & Bronze               & 0.40 \\
J0420.3-3745    & BCU  & (1.06 $\pm$ 0.77)$\times$10$^{-12}$      & \multicolumn{1}{c}{-} & IC190504A      & Bronze               & 0.39 \\
\hline
\end{tabular}

\label{tab:coincidence_table}
\end{table*}
\footnotetext[1]{Classification in 4FGL-DR3}
\footnotetext[2]{4FGL-DR3 Energy Flux from 100MeV to 100 GeV}
\footnotetext[3]{Redshift from \citealt{2019A&A...632A..77C}.}
\footnotetext[4]{Redshift from \citealt{2022MNRAS.510.2671P}.}

\subsection{Results and discussion}
\label{sec:results}
\begin{figure}
\sidecaption
    \includegraphics[width=\linewidth]{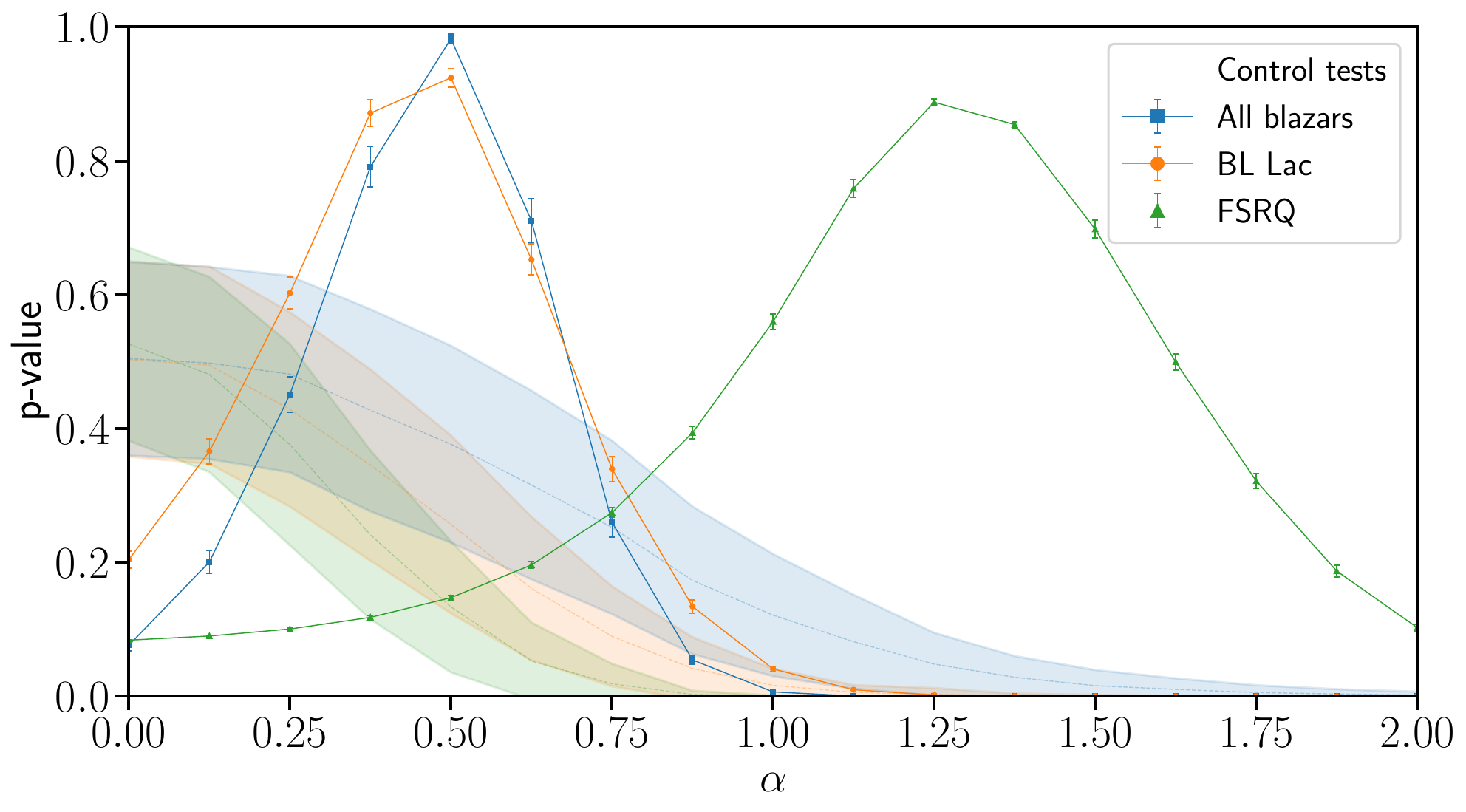}\\
    \includegraphics[width=\linewidth]{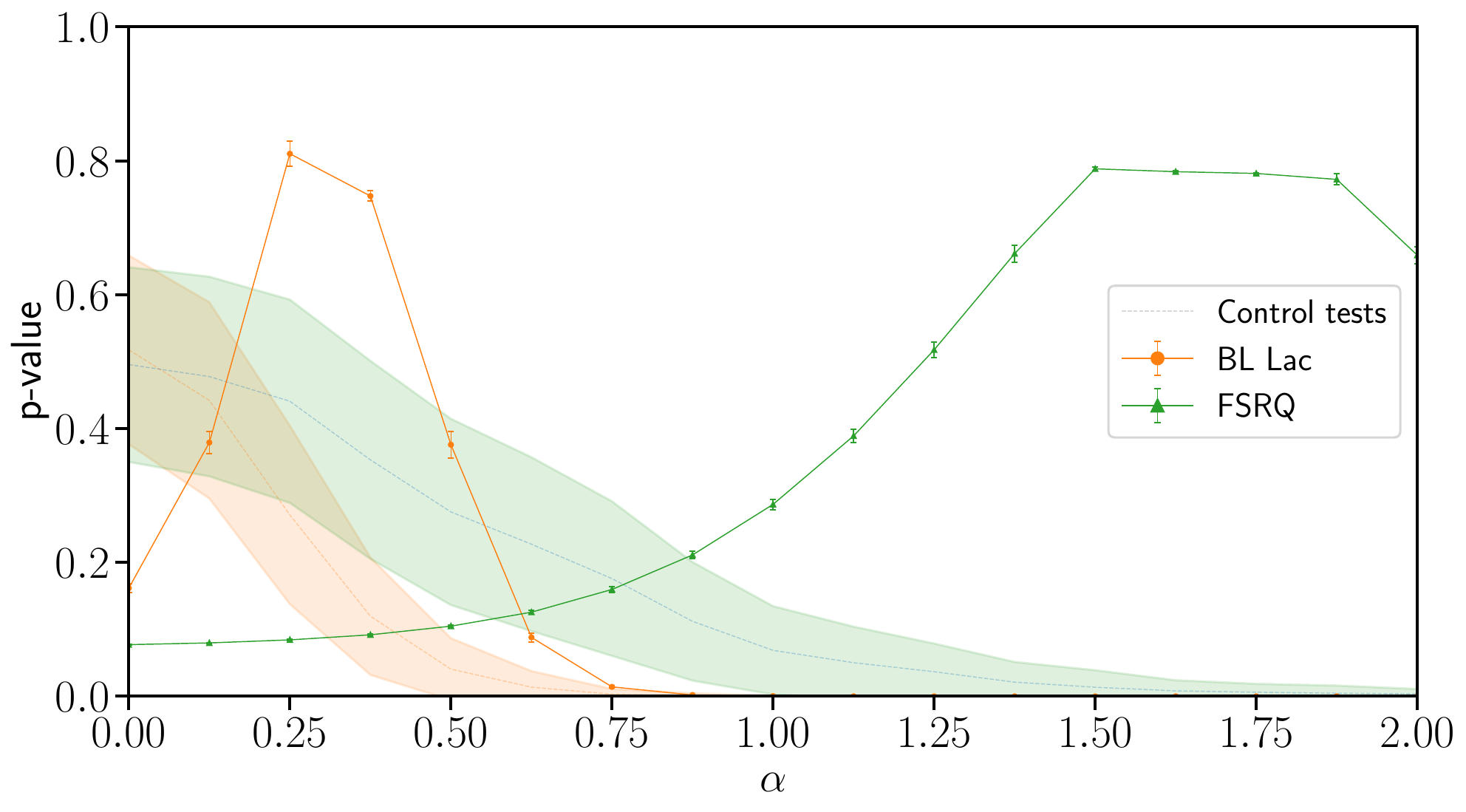}
        \caption{$p$-values from the K-S tests as a function of the index $\alpha$ from Eq. \ref{eq:lnu_lgamma}. The solid lines show the results for each class of sampled neutrino blazar candidates. The error bars show the standard deviation of each p-value after repeating the test 10$^{3}$ times for each $\alpha$. The dashed lines show the results for the test performed on the control samples with shaded areas indicating the 1$\sigma$ uncertainty regions from the N = 1000 Monte Carlo samples drawn for each $\alpha$ and source class. The upper panel shows the results from \textit{T1} (gamma-ray energy flux as proxy for the luminosity), while the lower panel shows the results from \textit{T2} (using only sources with known redshift).}
	\label{fig:pval_vs_power}
\end{figure}

Figure \ref{fig:pval_vs_power} shows the results of our \textit{K-S} tests. In the upper panel, we show the results of \textit{T1}, where we use the energy flux as a proxy for the luminosity. The curves of the p-value for each \textit{K-S} test as a function of $\alpha$ in the grid show two main features: there are two distinct behaviors between the subsamples of FSRQs and BL Lacs, and there is a single main peak for each p-value curve. We do not find a statistically significant excess at any value of $\alpha$. However, we find a trend with preferred $\alpha$ values in the range 1.2 $\leq \alpha \leq$ 1.5 for FSRQs ($p$ > 0.8, green curve) and in 0.25 $\leq \alpha \leq$ 0.65 for BL Lacs. For the FSRQ population, all other values of $\alpha$ are disfavored with p-values approaching $\leq$ 0.1 as $\alpha$  approaches 0 and 2.0. Similar behavior is observed for the BL Lacs where the p-values for higher values of $\alpha$ drop quickly to $p$ < 0.1 for $\alpha$ > 1.0. A moderate ($p \sim $ 0.2) is however still observed for the BL Lac sample for $\alpha \sim$ 0. This can be explained by the fact that the sample is still significantly contaminated by random coincidences, and this shifts the p-value curve to lower values of $\alpha$. Ultimately, the blue solid line shows the results for the whole sample (including blazars of uncertain type) and shows the highest correlation ($p$ > 0.95) for $\alpha \sim$ 0.5.

We compare our measurements to a control sample, and we show in the same plot the average p-values and the 1-$\sigma$ intervals as a function of $\alpha$ as dashed lines and shaded areas, respectively. The maximum p-values for the control samples occur at $\alpha$ = 0, as expected for a correlation with a random sample.

The lower panel of Figure \ref{fig:pval_vs_power} shows the results of \textit{T2} for the samples of the two optical classes of BL Lacs and FSRQs. The two samples exhibit similar behavior to the one observed using the energy flux in \textit{T1}. The highest p-value levels are lower than in \textit{T1}, likely because of the reduced sample that can be used in this test. A main difference for the FSRQs is that \textit{K-S} tests show high p-values ($p$ > 0.6) up to $\alpha$ = 2.0, while for the BL Lacs the highest correlation is more constrained around $\alpha$ = 0.25 ($p$ > 0.8).

The relation between the neutrino and gamma-ray luminosities in astrophysical sources is usually expressed in the form of Eq. \ref{eq:lnu_lgamma}, and often discussed in the literature. \citet{2016PhRvD..94j3006M} discuss the theoretical expectations of this functional dependence for different classes of cosmic neutrino sources in detail. For environments typical of FSRQs, the neutrino production is expected to be dominated by the interaction of high-energy protons with external photon fields originating from the broad line region, accretion disk, or dust torus. In \cite{2014PhRvD..90b3007M}, the photomeson production efficiency $f_{p\gamma}$ from optical and infrared data was found to be proportional to $L_{AD}^{1/2}$, where $L_{AD}$ is the accretion disk luminosity. Therefore, from the simple assumption that the cosmic-ray luminosity $L_{CR}$ is proportional to $L_{AD}$, one easily obtains 
\begin{equation}
 L_{\nu} \propto L_{CR} \cdot L_{AD}^{1/2} \propto L_{\gamma}^{3/2} .
\end{equation}
This implies that the theoretically predicted value for FSRQs would be $\alpha$ = 1.5. This is compatible with the observed excesses in both \textit{T1} and \textit{T2} for our sample of FSRQs neutrino candidates.\\
For BL Lac neutrino candidates, theoretical predictions are less constrained. Given the absence of external photon fields, the main target photon field arises from the synchrotron radiation present in the blazar jet; therefore, the luminosity dependence usually expected for these objects can be anything between $\alpha$ = 1 (the target photons from the low-frequency bump do not increase with the high-frequency bump) and $\alpha$ = 2 (target photons are proportional to $L_{\gamma}$, \citealt{2016PhRvD..94j3006M,2015MNRAS.451.1502T,2014PhRvD..90b3007M,2015MNRAS.448.2412P}).\\

\section{Conclusions}
\label{sec:conclusions}
In this paper, we present the \Fermi-LAT follow-up strategy for high-energy neutrino realtime alerts and the gamma-ray results from the past 7 years of activity in these searches. The spatial resolution of current neutrino detectors poses a strong limit on the searches for astrophysical counterparts, and the tentative associations with astrophysical sources suffer from high chance coincidence probabilities (see also \citealt{2022A&A...666A..36L}). With the increase in the number of realtime alerts, a growing sample of gamma-ray sources that are coincident with multiple neutrinos is emerging. Most of these coincidences are caused by very large positional uncertainties in the arrival directions of the neutrinos; therefore, detailed multiwavelength studies are necessary to select possible source candidates.

After the outstanding coincidence between the flaring gamma-ray blazar \TXS and the realtime event IC170922A \citep{IceCube:2018dnn}, the only realtime event pinpointing a single, powerful gamma-ray blazar is IC190730A which is coincident with \PKS \citep{2020ApJ...893..162F}. In this case, the source was not observed to be flaring at the time of the neutrino detection, and this motivated our further studies of the gamma-ray/neutrino connection that go beyond the temporal correlation between the two messengers. 

In the current sample of blazars that are coincident with well-reconstructed neutrino alerts, we do not find a significant difference in terms of gamma-ray brightness for the neutrino blazar candidates. However, we identify a trend towards brighter sources, best described by a L$_{\nu}$ $\propto$ L$_{\gamma}^{\alpha}$ relation, with 1.2 $\leq \alpha \leq$ 1.5 and 0.25 $\leq \alpha \leq$ 0.65 for FSRQs and BL Lacs, respectively.
The differences in the relation between the two sub-populations of blazars is expected from theoretical models because of the different characteristics of their environments. In the future, a larger sample of neutrino blazar candidates will refine the tests of these correlations and better constrain the degree of correlation between neutrino production and gamma-ray brightness. Although we restricted our study to a sample of well-reconstructed alerts, the sample is likely contaminated by a significant fraction of random coincidences.\\
The \Fermi-LAT follow-up program for realtime neutrino alerts is ongoing, and we are continously improving our strategies and synergies with multiwavelength facilities. Follow-up observations of these realtime alerts with the \Fermi-LAT is crucial for the prompt identification of interesting neutrino source candidates.

\begin{acknowledgements}
The authors would like to thank Cristina Lagunas Gualda, Xavier Rodrigues, Anastasiia Omeliukh, Deirdre Horan, Melissa Pesce-Rollins, Jean Ballet and C.C. Teddy Cheung for the helpful comments and discussions.\\ 
AF acknowledges the support from the DFG via the Collaborative Research Center SFB1491 \textit{Cosmic Interacting Matters - From Source to Signal}. SG and AF acknowledge support by the Initiative and Networking Fund of the Helmholtz Association through the Young Investigator Group program. Part of this work was supported by the German \emph{Deut\-sche For\-schungs\-ge\-mein\-schaft, DFG\/} project number Ts~17/2--1.\\
GP acknowledges the support by ICSC – CentroNazionale di Ricerca
in High Performance Computing, Big Data and Quantum Computing,
funded by European Union – NextGenerationEU.\\
This work was supported by the European Research Council, ERC Starting grant \emph{MessMapp}, S.B. Principal Investigator, under contract no. 949555.\\
CB acknowledges that this paper has been conducted during and with the support of the Italian national inter-university PhD programme in Space Science and Technology.\\

The \textit{Fermi} LAT Collaboration acknowledges generous ongoing support
from a number of agencies and institutes that have supported both the
development and the operation of the LAT as well as scientific data analysis.
These include the National Aeronautics and Space Administration and the
Department of Energy in the United States, the Commissariat \`a l'Energie Atomique
and the Centre National de la Recherche Scientifique / Institut National de Physique
Nucl\'eaire et de Physique des Particules in France, the Agenzia Spaziale Italiana
and the Istituto Nazionale di Fisica Nucleare in Italy, the Ministry of Education,
Culture, Sports, Science and Technology (MEXT), High Energy Accelerator Research
Organization (KEK) and Japan Aerospace Exploration Agency (JAXA) in Japan, and
the K.~A.~Wallenberg Foundation, the Swedish Research Council and the
Swedish National Space Board in Sweden.
 
Additional support for science analysis during the operations phase is gratefully
acknowledged from the Istituto Nazionale di Astrofisica in Italy and the Centre
National d'\'Etudes Spatiales in France. This work performed in part under DOE
Contract DE-AC02-76SF00515.\\
This work made use of Astropy:\footnote{http://www.astropy.org} a community-developed core Python package and an ecosystem of tools and resources for astronomy \citep{astropy:2013, astropy:2018, astropy:2022}.

\end{acknowledgements}

\bibliographystyle{aa} 
\bibliography{biblio} 
\end{document}